\documentclass{article}

\usepackage{arxiv}

\usepackage[utf8]{inputenc} 
\usepackage[T1]{fontenc}    
\usepackage{hyperref}       
\usepackage{url}            
\usepackage{booktabs}       
\usepackage{amsfonts}       
\usepackage{amsmath} 
\usepackage{amsthm} 
\usepackage{mathtools}
\usepackage{amssymb}
\usepackage{nicefrac}       
\usepackage{microtype}      
\usepackage{lipsum}
\usepackage{graphicx}
\graphicspath{ {./Img/} }

\theoremstyle{plain}

\theoremstyle{definition}

\newtheorem{remark}{Remark}
\newtheorem{example}{Example}

\title{The application of theory of probability to the modelling of chemical kinetics systems}

\author{
 Maxim Nazarov\\
  Chair of Higher Math 1\\
  Moscow Institute of Electronic Technology\\
  \texttt{Nazarov-Maximilian@yandex.ru} \\
}

\begin{document}
\maketitle
\begin{abstract}
We consider a model of chemical kinetics for which the derivation of equations does not rely on the law of mass action, but is rather based on such principles as the joint probability and the geometric probability. For this model a generalization is constructed for the case of reaction-diffusion systems in heterogeneous medium  with respect to the convective and diffusive transfer of heat. The construction of this generalization is carried out by an alternative methodology  which is based fully on a systems of ordinary differential equations, without a transition to the partial derivatives. The description of this new method is a bit similar  to the finite volume method, except that it uses statistical simplifying positions and geometric probability to describe the diffusion processes. Such approach allows us to greatly simplify the numerical implementation of the resulting model, as well as to simplify the quantitative analysis of it with dynamical systems theory. Moreover, the efficiency of the parallel implementation of the numerical method is increased for the resulting model. In addition, we will consider an application of this model for the description of some example reaction with quasi-periodic regime, as well as consider an algorithm for the transition from standard models with dimensional kinetic constants to its formalism.
\end{abstract}


\section*{Introduction}

The most often used models to describe chemical kinetics are those, which are based on differential equations with dimensional kinetic constants (see examples in \cite{Kenneth1991} and \cite{Emanuel1984}). These models are also used as the basis for defining reaction-diffusion systems with partial differential equations: $\partial{u}_{i}/\partial t = D\cdot \nabla^{2} u_{i} + f$ (for more details, see \cite{Aris1975}). It should be noted that this approach is the simplest in terms of formalization, but is far from optimal in practice.
The main drawback of models with dimensional constants is in their lack of physical rigour. Furthermore, the difference in the dimensions of kinetic constants for systems with a large number of reactions significantly complicates comparative kinetic analysis. The main drawbacks of reaction-diffusion models based on partial derivatives are the difficulty of obtaining a numerical solution and the difficulty of qualitatively analysing this model using methods from dynamical systems theory.

As part of the work by  \cite{Nazarov_Chem1}, a model of chemical kinetics was proposed that aims to provide the most complete and physically rigorous description of kinetics.
To achieve this goal  this model was constructed exclusively for a systems of elementary reactions.
In turn, to construct a more rigorous model for elementary reactions in \cite{Nazarov_Chem1}  the methods of theory of probability  were used, such as the probability of a complex event and the principle of geometric probability.
Subsequently, within the framework of the work \cite{Nazarov_Chem3} this model was refined, and a method for organizing automated analysis of elementary reactions  was also incorporated.

To describe reaction-diffusion systems within the framework of \cite{Nazarov2011}  a special method was proposed that allows modelling such systems using only ordinary differential equations (without switching to partial derivatives). 
This method uses geometric probability as its basis, as well as a special scheme for stabilizing calculations and compensating for overflowing of spatial cells. 
This method was subsequently generalized to the case of heat convection and diffusion in the article \cite{Nazarov_Chem2}, and its accuracy was further assessed in the work \cite{Nazarov2014}.

In this paper, we will refine the model from \cite{Nazarov_Chem2}, taking into account the latest version of the methodology from \cite{Nazarov2014} and using as a basis the new basic model of chemical kinetics from \cite{Nazarov_Chem3}. We will also consider specific examples of its application to the description and analysis of complex chemical kinetic systems.


\section{Description of the chemical kinetics model}
\label{sec:Description}
To describe the model space  we will use a formalism similar to the description of the space in \cite{Nazarov_Chem2}. The model space will be a parallelepiped of volume $ V = h \cdot l \cdot w$ with matched side lengths $ h=a \cdot H(a) $, $ l= a \cdot L(a) $, $ w= a \cdot W(a) $. Precisely by analogy with the model in \cite{Nazarov_Chem2}  this space can be uniquely divided into cubic cells of volume $ a^3 $. 
To work with these cells we will use the following notation:
\begin{enumerate}
	\item $ d \in \left\lbrace \uparrow \, , \downarrow \,
	, \rightarrow, \leftarrow, \odot, \otimes \right\rbrace  $~--- direction of diffusion;
	\item $ \overline{d} $~--- opposite direction for $ d $;
	\item $ d(z,x,y) $~--- neighbouring cell with $ (z,x,y) $ in the direction $ d $.
\end{enumerate}

To apply the scheme from \cite{Nazarov2014}  we need to introduce a reference time $t_{r}$ and ${r}_{p}(t_{r})$~-- the maximum migration radii for molecules $p$ during the reference time $t_{r}$.
Using the description for Brownian motion  we can estimate this radius to a first approximation as ${r}_{p}(t_{r}) = \sqrt{2\cdot D_{p} \cdot t_{r}}$ (for $t_{r}\rightarrow 0$), where $D_{p}$~-- is the diffusion coefficient. We denote the maximum radius $r_{p}$ as $\Delta l = \max\limits_{p} r_{p}$.
Since the model allows for temperature changes in cells, then the coefficient $D_{p}$ will be a function that depends on temperature:
\begin{gather*}
	D_{p}(T) = \dfrac{D_{p}(T_{E})\cdot T}{T_{E}} \cdot \dfrac{\mu_{p}(T_{E})}{\mu_{p}(T)}
\end{gather*}
In this expression  $T_{E}$~--- is the reference temperature for which all substances under consideration are in a liquid state of aggregation  and $\mu_{p}(T)$~--- is the dynamic viscosity of the substance $p$.

In exact analogy with the model from the article \cite{Nazarov_Chem2}  we will denote the set of all reactions of the reaction system under study as $\mathcal{R}$, where an individual reaction $ r \in \mathcal{R}$ has the following form:
\begin{gather*} \label{nazarov:Chem_Kinetics:Reaction_Schema_eq1}
	k^{+}_{a_{1}} a_{1} + \ldots
 + k^{+}_{a_{n}} a_{n}  \,
			\xrightleftharpoons[v^{-}, \, E^{-}]{v^{+}, \, E^{+}}  \,
		k^{-}_{b_{1}}  b_{1} + \ldots
 + k^{-}_{b_{p}}  b_{p}
\end{gather*} 
The set of substrates of reaction $r$ we denote as $A(r)=\left\lbrace a_{1}, \ldots \, , a_{n(r)} \right\rbrace$  and the set of products as $B(r)=\left\lbrace b_{1}, \ldots \, , b_{p(r)} \right\rbrace$. Moreover, the set of all molecules of the reaction system, by analogy with the model from \cite{Nazarov_Chem2}, we will divide into two subsets: independent reactants $S$ and intermediate products $I$.
In turn, the variables in the generalized model will be:
\begin{itemize}
	\item $N_{p}^{(z,x,y)}(t)$ --- a number of molecules uniformly mixed in the cell $(z,x,y)$ at time $t$;
	\item $\mathcal{\hat{N}}_{p}^{(z,x,y)}(t)$ --- a number of  molecules that entered the boundary regions of the cell at time $t$;
	\item  $T^{(z,x,y)}(t)$ --- the temperature in the cell at time $t$.
\end{itemize}
We will describe the dynamics of the generalized model using a system of differential equations:
\begin{gather} \label{nazarov:Chem_Kinetics_gener:main_system_eq1}
	  \dfrac{\mathrm{d}N_{p}^{(z,x,y)}}{\mathrm{d}t} = 
		F^{(z,x,y)}_{p} + \mathrm{Dif}^{(z,x,y)}_{p (-)} + \beta_{p}	 \cdot \mathcal{\hat{N}}_{p}^{(z,x,y)}; \quad
	\dfrac{\mathrm{d}\mathcal{\hat{N}}_{p}^{(z,x,y)}}{\mathrm{d}t} = \mathcal{F}^{(z,x,y)}_{p} +
		\mathrm{Dif}^{(z,x,y)}_{p (+)} - \beta_{p}	 \cdot \mathcal{\hat{N}}_{p}^{(z,x,y)}		
\end{gather}
\begin{gather} \label{nazarov:Chem_Kinetics_gener:main_system_eq2}
	 \dfrac{\mathrm{d}T^{(z,x,y)}}{\mathrm{d}t} =  
		\mathrm{Reac}^{(z,x,y)}_{T} + \mathrm{Con}^{(z,x,y)}_{T} +   \mathrm{Dif}^{(z,x,y)}_{T}
\end{gather}
Let us denote the number of reactions $r$ occurring in parallel per unit time between mixed reactants $N_{p}^{(z,x,y)}(t)$ as $\omega^{+}_{r}(z,x,y)$  and the number of reactions inverse to them as $\omega^{-}_{r}(z,x,y)$. For molecules in the boundary regions $\mathcal{\hat{N}}_{p}^{(z,x,y)}(t)$ we will use similar notations $\hat{\omega}^{+}_{r}(z,x,y)$ and $\hat{\omega}^{-}_{r}(z,x,y)$.
Given the notation we have introduced, we can write the following expressions for the gain and loss of $N_{p}^{(z,x,y)}$ molecules during chemical reactions:
\begin{align} \label{nazarov:Chem_Kinetics_gener:loss_surplus_eq2}
	\nonumber
	F^{(z,x,y)}_{p} =  & \sum_{\substack{r \, \in \, \mathcal{R}	  \, : \\  \, p \in A(r)}}
		 {k^{+}_{p}(r)}\cdot \omega^{-}_{r}(z,x,y) + 
				\sum_{\substack{r \, \in \, \mathcal{R} \, : \\ p \in B(r)}} {k^{-}_{p}(r)}\cdot \omega^{+}_{r}(z,x,y) -	 \\
	 	- & \sum_{\substack{r \, \in \, \mathcal{R}
	 \, : \\ p \in A(r)}} 
		{k^{+}_{p}(r)} \cdot \omega^{+}_{r} (z,x,y)
			 -	\sum_{\substack{r \, \in \, \mathcal{R}
			 \, : \\ p \in B(r)}} {k^{-}_{p}(r)} \cdot \omega^{-}_{r}(z,x,y) 
\end{align}
The expression for $\mathcal{F}^{(z,x,y)}_{p}$ will look completely similar, with the only exception that instead of $\omega^{+}_{r}(z,x,y)$ and $\omega^{-}_{r}(z,x,y)$ we will use $\hat{\omega}^{+}_{r}(z,x,y)$ and $\hat{\omega}^{-}_{r}(z,x,y)$. These quantities can be found using the following\footnote{For $\hat{\omega}^{+}_{r}$ and $\hat{\omega}^{-}_{r}$  we need to replace $N_{p}$ with $\mathcal{\hat{N}}_{p}$  and the volume of the cell $a^{3}$ with the volume of the boundary zone $a^{3} - (a-2\Delta l)^{3}$.}  formulas:
\begin{gather} \label{nazarov:Chem_Kinetics_gener:omega_eq4+}
					\omega^{+}_{r}(z,x,y) = v^{+}(r)\cdot
									 e^{\frac{-E^{+}(r)}{R\cdot T}} \cdot
									  \prod_{j=1}^{n(r)} \left( \dfrac{N^{(z,x,y)}_{a_{j}}}{a^{3}\cdot\rho_{a_{j}}^{N}(T)} \right)^{k_{a_{j}}^{+}(r)}\! \! \!\! \! \!\!\! \! \! \cdot \min_{j}\left( \dfrac{N^{(z,x,y)}_{a_{j}}}{k_{a_{j}}}\right)
\end{gather}
\begin{gather} \label{nazarov:Chem_Kinetics_gener:omega_eq4-}
					\omega^{-}_{r}(z,x,y) = v^{-}(r)\cdot
									 e^{\frac{-E^{-}(r)}{R\cdot T}} \cdot
									  \prod_{j=1}^{p(r)} \left( \dfrac{N^{(z,x,y)}_{b_{j}}}{a^{3}\cdot\rho_{b_{j}}^{N}(T)} \right)^{k_{b_{j}}^{-}(r)}\! \! \!\! \! \!\!\! \! \! \cdot \min_{j}\left( \dfrac{N^{(z,x,y)}_{b_{j}}}{k_{b_{j}}}\right)
\end{gather}
This expression is derived using elementary considerations of probability theory:
\begin{enumerate}
	\item $\prod\limits_{j=1}^{n(r)} \left( \dfrac{N^{(z,x,y)}_{p}}{a^{3}\cdot\rho_{p}^{N}(T)} \right)^{k_{p}(r)}$~--- estimate of the probability of reactant collision;
	\item $\exp\left( {\dfrac{-E(r)}{R\cdot T}}\right) $~--- estimate of the probability of reaction initiation upon collision of reactants;
	\item $\min\limits_{j}\left( \dfrac{N^{(z,x,y)}_{p}}{k_{p}}\right)$~--- the maximum possible number of reactions per unit of time;
	\item $v(r)$~--- elementary reaction rate (provided that it has already been initiated).
\end{enumerate}
 
If we introduce the total heat capacity of the cell $C^{(z,x,y)}$, then we can compose the following expression to calculate the magnitude of the temperature change during the reactions $ \mathrm{Reac}^{(z,x,y)}_{T}$:
\begin{gather} \label{nazarov:Chem_Kinetics_gener:T_new_eq7}
	 \mathrm{Reac}^{(z,x,y)}_{T}= 
		\dfrac{1}{C^{(z,x,y)}} \sum_{r \in R} \left( E^{-}_{r} - E^{+}_{r} \right)  
			\left( \omega^{+}_{r}(z,x,y) -	\omega^{-}_{r}(z,x,y) \right)
\end{gather}
Since the model takes into account all possible intermediate products $i \in I$ that can arise in the reaction system, then the heat capacity $C$ of an individual cell $(z,x,y)$ can be calculated\footnote{In the formula \eqref{nazarov:Chem_Kinetics:C_t_eq8} $c^{N}_{p}(T)$ --- are the molecular heat capacities as functions of temperature $T$. The molecular heat capacity can be expressed in terms of the molar heat capacity as $c^{N}_{p}(T) = c^{\nu}_{p}(T) N_{A}$, where $N_{A}$~--- is the Avogadro's number.}  according to the additive law:
\begin{gather} \label{nazarov:Chem_Kinetics:C_t_eq8}
	C^{(z,x,y)} = \sum\limits_{s \in S} c^{N}_{s}(T)\cdot N^{(z,x,y)}_{s} + 
	\sum\limits_{i \in I} c^{N}_{i}(T)\cdot N^{(z,x,y)}_{i} 
\end{gather}


To calculate the diffusion balance of molecules $p$ in a separate cell $(z,x,y)$  it is necessary to take into account the flows in all permissible diffusion directions:
\begin{gather} \label{nazarov:Diff_eq3}
	\mathrm{Dif}^{(z,x,y)}_{p(+)} = \sum_{d} 
		\xi\left( \mathrm{Emig}_{\, p, \, \overline{d}}^{d(z,x,y)} - 
			\mathrm{Emig}_{\, p,\, d}^{(z,x,y)}\right)    
\end{gather}
\begin{gather} \label{nazarov:Diff_eq3-2}
  \mathrm{Dif}^{(z,x,y)}_{p(-)} = \sum_{d} 
		\mu\left( \mathrm{Emig}_{\, p, \, \overline{d}}^{d(z,x,y)} - 
			\mathrm{Emig}_{\, p,\, d}^{(z,x,y)}\right)
\end{gather}
In the formulas \eqref{nazarov:Diff_eq3} and \eqref{nazarov:Diff_eq3-2} the functions $\xi(x) = \sigma(x)\cdot x$ and $\mu(x) = \sigma(-x)\cdot x$ are defined either with the Heaviside function $\sigma(x)$ or the sigmoid function $\sigma^{c}(x) = {1}/({1 + e^{-x/c}})$, where $c\rightarrow 0$ and $c>0$. 

The flow $\mathrm{Emig}_{p,\, d}^{(z,x,y)}(t)$ of molecules $p$ from the cell $ (z,x,y) $ in the direction $ d $ can be calculated using the following formula (for the derivation of the formula  see the work \cite{{Nazarov2011}}):
\begin{gather} \label{nazarov:Emig_definition_eq4}
	\mathrm{Emig}_{\, p,\, d}^{(z,x,y)} = \dfrac{3 \cdot N^{(z,x,y)}_{p} \cdot r_{p} }{16\cdot t_{r}\cdot a} +
		 \dfrac{\Delta_{(z,x,y)}^{d, p} \cdot \rho_{p}^{N}(T) }{2\cdot  M^{\, d}_{(z,x,y)} \cdot t_{r}}	
\end{gather}
The quantity $M^{\, d}_{(z,x,y)} $ in the formula \eqref{nazarov:Emig_definition_eq4} is equal to the number of different types of molecules that at time $t$ are located in one of two cells: either in $(z,x,y)$ or in the neighbouring cell $d(z,x,y)$: 
\begin{gather*} 
	M^{\, d}_{(z,x,y)} = \begin{cases} 
		 1, & \mathrm{if} \quad  N^{(z,x,y)}_{i}=N^{d(z,x,y)}_{i}=0 \, \, (\forall i); \\
		\left|  \left\lbrace i \, | \, \, N^{(z,x,y)}_{i}>0 \vee N^{d(z,x,y)}_{i}>0 \right\rbrace \right| , & \mathrm{if}	\quad \mathrm{else}.
		\end{cases}
\end{gather*}

The model assumes that the exchange of matter occurs between the boundary regions of cells whose linear size is $\Delta l$.
To estimate the heat diffusion $\mathrm{Dif}_{T}^{(z,x,y)}$  we use the integral form of the Fourier law of thermal conductivity (see the derivation in \cite{Nazarov_Chem2}):
\begin{gather} \label{Heat_Diffusion}
	  \mathrm{Dif}^{(z,x,y)}_{T}   = 
	\dfrac{a^{2} \cdot \chi^{(z,x,y)} } {\Delta l \cdot C^{(z,x,y)} }\cdot  
	    \sum_{d} \chi^{d(z,x,y)} \dfrac{  T^{d(z,x,y)} - T^{(z,x,y)} }{   \chi^{d(z,x,y)} + \chi^{(z,x,y)}}
\end{gather}
In the equation \eqref{Heat_Diffusion}  the quantity $C^{(z,x,y)}$~--- is the heat capacity of the cell $(z,x,y)$  and the coefficient $\chi^{(z,x,y)}$ specifies the average thermal conductivity of the cell $(z,x,y)$, which can be roughly estimated using the weighted averaging formula as:
\begin{gather} \label{Chi_coef_Diffusion}
	\chi^{(z,x,y)} = \dfrac{\sum\limits_{p} \chi_{p}(T)   \cdot N^{(z,x,y)}_{p} / \rho_{p}(T) }{\sum\limits_{p }  N^{(z,x,y)}_{p} / \rho_{p}(T)}
\end{gather}
In the formula \eqref{Chi_coef_Diffusion} the coefficients $\chi_{p}(T)$~--- are the thermal conductivity coefficients for the given substances, which are assumed to have known functional dependences on temperature.

The expression for heat convection can be obtained using the coefficients $\mathrm{Emig}$, which provide an estimate of the diffusion balance of the substance (for a detailed derivation  see the work \cite{Nazarov_Chem2}):
 \begin{gather} \label{Heat_Convection}
	   \mathrm{Con}_{T}^{(z,x,y)}  =  
	\dfrac{ 1  } {C^{(z,x,y)}} 
	\cdot \sum\limits_{m} {c^{N}_{m}} \cdot  \sum\limits_{d} \left(  T^{d(z,x,y)} - T^{(z,x,y)} \right) \cdot  
	   \left(\mathrm{Emig}_{\, m, \, \overline{d}}^{d(z,x,y)} - \mathrm{Emig}_{\, m,\, d}^{(z,x,y)} \right) 
\end{gather}


The coefficient $ \Delta_{(z,x,y)}^{d, p} $ in the formula \eqref{nazarov:Emig_definition_eq4} gives an estimate of the relative filling of the boundary regions of the cells $(z,x,y)$ and $d(z,x,y)$ by $p$ molecules separately from other types of molecules:
\begin{gather} \label{nazarov:Delta_i}
	\Delta_{(z,x,y)}^{d, i} = \dfrac{\Delta_{(z,x,y)}^{d}
	 \cdot \mathcal{\hat{N}}^{(z,x,y)}_{i}}{\sum\limits_{\mathcal{\hat{N}}_{p}>0} \mathcal{\hat{N}}^{(z,x,y)}_{p}}
\end{gather}

The coefficient $ \Delta_{(z,x,y)}^{d} $ in the formula \eqref{nazarov:Delta_i} is introduced\footnote{As $\mathrm{sign}$  either the sign function or its approximation $ \mathrm{sign}_{c}(x)=(2/\pi)\cdot \arctan \left( x/c\right) $ is used.} to determine the total filling of the boundary regions of cells $(z,x,y)$ and $d(z,x,y)$ in excess of the permissible value ($\Delta l \cdot a^{2}$):  
\begin{gather} \label{nazarov:Delta_compensation}
	\Delta_{(z,x,y)}^{d} =  \xi \left(  \max\left\lbrace L_{(z,x,y)}^{d}, 
	 L_{d(z,x,y)}^{\overline{d}} \right\rbrace - \Delta l \cdot a^{2} \right)
	 \cdot
	  \mathrm{sign} \left( L_{(z,x,y)}^{d} -  L_{d(z,x,y)}^{\overline{d}}
	 					\right)  
\end{gather}

\begin{remark}
	Note that the coefficient $ \Delta_{(z,x,y)}^{d} $ is introduced to stabilize the model's predictions under harsh conditions: large diffusion coefficients, active heat release during reactions, and so on. In most cases  the coefficient $ \Delta_{(z,x,y)}^{d} $ can be forced to zero, effectively eliminating the second term from equation \eqref{nazarov:Emig_definition_eq4}.
\end{remark}

The value $ L_{(z,x,y)}^{d} $ from the formula \eqref{nazarov:Delta_compensation} specifies the volume that the molecules in the boundary region of the cell $ (z,x,y) $ should occupy after the exchange with the neighbouring cell in the direction $ d $:
\begin{gather} \label{nazarov:L_zxy_definition}
	L_{(z,x,y)}^{d} = V_{L}^{(z,x,y)} -
						 \delta V_{L}^{(z,x,y)} +
						 \delta V_{L}^{d(z,x,y)}
\end{gather} 
The value $ V_{L}^{(z,x,y)} $ from \eqref{nazarov:L_zxy_definition} is the current\footnote{$k(z,x,y)$ is the number of boundary zones of the cell $(z,x,y)$. } filling of the boundary region of the cell $(z,x,y)$:
\begin{gather} \label{nazarov:V_L_definition}
	V_{L}^{(z,x,y)} = \sum_{p} \mathcal{\hat{N}}^{(z,x,y)}_{p}  
	 / \big( \rho_{p}(T) \cdot k(z,x,y) \big)	 
\end{gather} 
The quantity $ \delta V_{L}^{(z,x,y)}(t) $ is the potential total increase in the volume of molecules from the boundary zone:
\begin{gather} \label{nazarov:Delta_V_L_definition}
	\delta V_{L}^{(z,x,y)} = \delta V_{F}^{(z,x,y)}(t) + \delta V_{M}^{(z,x,y)}(t)
\end{gather} 
The term $\delta V{F}$ is responsible for the increase in volume as a result of reactions in the boundary zone:
\begin{gather} \label{nazarov:Delta_V_F_definition}
	\delta V_{F}^{(z,x,y)} = \sum_{p}   \mathcal{F}^{(z,x,y)}_{p} / \big( \rho_{p}(T) \cdot k(z,x,y) \big)
\end{gather} 
The second term $\delta V_{M}^{(z,x,y)}(t)$ in the formula \eqref{nazarov:Delta_V_L_definition} is responsible for the increase in volume as a result of the exchange of molecules with the boundary zone of the neighbouring cell $d(z,x,y)$:
\begin{gather} \label{nazarov:Delta_V_L_M_definition}
	\delta V_{M}^{(z,x,y)}(t) = \sum_{p} \dfrac{3   }{16 }\cdot N^{(z,x,y)}_{p}  \cdot    \dfrac{ r_{p}}{a\cdot \rho_{p}(T)} 
\end{gather} 

To estimate the parameter $\beta_{p}$ in the equations  \eqref{nazarov:Chem_Kinetics_gener:main_system_eq1} and  \eqref{nazarov:Chem_Kinetics_gener:main_system_eq2} we introduce the simplifying provision that the molecules that have shifted by a distance $a/2$ within the cell can be considered to be dispersed in this cell. 
Using the constraint equation $r = \sqrt{2 D t_{s_{p}}}$  we obtain an estimate for the time $t_{s_{p}} = a^{2} / (4 D)$. For definiteness, let $(1-c)$ percent of the original $\mathcal{\hat{N}}_{i}(0)$ dissolve during this time, where $0<c\ll1$.
Considering the case of only internal diffusion within a single cell  we can write the following expression for the numbers $\mathcal{\hat{N}}_{p}(t)$:
\[
	\mathcal{\hat{N}}_{p}(t) = \mathcal{\hat{N}}_{p}(0) \cdot e^{-\beta_{p} t}
\]
Substituting into this formula our data $\mathcal{\hat{N}}_{p}(t_{s_{p}}) = c \cdot \mathcal{\hat{N}}_{p}(0)$ and $t_{s_{p}} = a^{2} / (4 D)$  we obtain as a result an estimate for the parameter $\beta_{p}$:
\begin{gather} \label{nazarov:mad_coefficient}
	 \beta_{p} = \dfrac{4 D}{ a^{2}} \log \left( c^{-1} \right) =  \log \left( c^{-\frac{4 D}{ a^{2}}} \right)
\end{gather}
 

\section{Practical application of kinetics model}

In practice  among all existing models of chemical kinetics, the models with dimensional constants are the most often used (see examples \cite{Kenneth1991} and \cite{Emanuel1984}). We will consider the relationship between the \eqref{nazarov:Chem_Kinetics_gener:main_system_eq1}--\eqref{nazarov:Delta_V_L_M_definition} model and the model with dimensional constants based on a two examples of elementary reactions.

\begin{example} \label{nazarov:Chem_Kinetics_2:example1}
Let us consider the irreversible reaction of $B \rightarrow C$ transformation at $T=const$.\\
If the molar concentrations $[p](t) = \nu_{p}(t)/V$ are used as variables, then we can write the following equation for the model with dimensional constant $k$ for this idealized reaction:
\[
	\dfrac{\mathrm{d}[C] }{\mathrm{d}t} =  k \cdot [B] \cdot e^{\left(  -\frac{E^{+}}{R\cdot T} \right)}, \quad  [k] =\left[\frac{1}{sec}\right]
\]
Using the fact that the number of molecules is equal to $N_{C}(t)=\nu_{C}(t)\cdot N_{A}=[C](t)\cdot N_{A}\cdot V$ and denoting the molar density of substance $B$ as $\rho^{\nu}_{B}$, then we can write the following equation for this reaction in accordance with the model \eqref{nazarov:Chem_Kinetics_gener:main_system_eq1}--\eqref{nazarov:Delta_V_L_M_definition} (we assume $H=L=W=1$):
\[
	\dfrac{\mathrm{d}[C] }{\mathrm{d}t} =  v^{+} \cdot  \frac{[B]^{2}}{\rho^{\nu}_{B}} \cdot e^{\left(  -\frac{E^{+}}{R\cdot T} \right)}
\]
As a result  we obtain the following equation for the constant $k$ and the reaction rate $v^{+}$:
\begin{gather*} \label{nazarov:Chem_Kinetics_2:v_and_k_connection}
	v^{+} = {k\cdot  \rho^{\nu}_{B}} / [B]
\end{gather*}
By definition, the kinetic constant $k$ is introduced for the case of unit molar concentrations of reactants ($[B]=1$). As a result  we obtain the relationship: $v^{+} = {k\cdot \rho^{\nu}_{B}}$.
\end{example}

A specific reaction that will be close to the idealized scheme of Example \label{nazarov:Chem_Kinetics_2:example1} is the dissociation reaction of hydrogen peroxide $H_{2}O_{2} \rightarrow 2 OH$, provided that the reaction occurs in the gas phase. Data on the dimensional constant ($k=10^{13}$ 1/c) for this reaction are given in \cite{Emanuel1984} on page 139. The density of hydrogen peroxide for the liquid phase is: $\rho_{H_{2}O_{2}} = 1.44 \cdot 10^{-9}$ g/m$^{3}$, and the molecular mass is correspondingly $M_{H_{2}O_{2}} = 34.01 $ g/mol. 
 \[
	H_{2}O_{2} \rightarrow 2 OH: \quad \quad v^{+} \approx \dfrac{k\cdot \rho_{H_{2}O_{2}}}{M_{H_{2}O_{2}}} \approx 4.223 \cdot 10^{2}  \dfrac{1}{c}
\]

\begin{example}
Consider the irreversible synthesis reaction $B + C \rightarrow D$ for the isothermal approximation $T=const$. If molar concentrations are used as variables, then we can write the following equations for the model with dimensional constant $k$:
\[
	\dfrac{\mathrm{d}[D] }{\mathrm{d}t} =  k \cdot [B] \cdot [C] \cdot e^{\left(  -\frac{E^{+}}{R\cdot T} \right)}, \quad  [k] =\left[\frac{m^{3}}{mol\cdot sec}\right]
\]
We will also construct the equations of the  \eqref{nazarov:Chem_Kinetics_gener:main_system_eq1}--\eqref{nazarov:Delta_V_L_M_definition} model for this reaction for the case $H=L=W=1$:
\[
 \dfrac{\mathrm{d}[D] }{\mathrm{d}t}  =  v^{+} \cdot  \left(\frac{[B]}{\rho^{\nu}_{B}}\right) \cdot \left(\frac{[C]}{\rho^{\nu}_{C}}\right) \cdot e^{\left(  -\frac{E^{+}}{R\cdot T } \right)} \cdot \min ([B], [C])
\]
As a result, we obtain a constraint equation for the constant $k$ and the reaction rate $v^{+}$ of the form:
\begin{gather}  \label{nazarov:Chem_Kinetics_2:v_and_k_connection3}
	v^{+} = k\cdot  \rho^{\nu}_{B}\cdot \rho^{\nu}_{C} / \min ([B], [C])
\end{gather}
By definition, the kinetic constant $k$ is introduced for the case of unit molar concentrations of reactants ($[B]=[C]=1$). As a result  we obtain the relationship: $v^{+} = {k\cdot \rho^{\nu}_{B}\cdot \rho^{\nu}_{C}}$.
\end{example} 
A specific reaction that will be close to the idealized scheme of example 2 is the recombination reaction of butane $C_{2}H_{5} + C_{2}H_{5}\rightarrow C_{4}H_{10}$ in the gas phase. In the work \cite{Emanuel1984} on page 138, data are given for this reaction on the size constant $k=3.5\cdot 10^{11}$ mol/s, and the molecular mass of the ethyl group $C_{2}H_{5}$ is $M_{C_{2}H_{5}} = 29.05 $ g/mol. 
Considering that the density of the ethyl group $C_{2}H_{5}$ in the liquid phase can be roughly estimated as $\rho_{C_{2}H_{5}} \approx 0.56 \cdot 10^{-9}$ g/m$^{3}$, we obtain:
\[
	C_{2}H_{5} + C_{2}H_{5}\rightarrow C_{4}H_{10}: \quad \quad v^{+} \approx  k\cdot   \left(\dfrac{\rho_{C_{2}H_{5}}}{M_{C_{2}H_{5}}}\right)^{2}  \approx 13 \cdot 10^{-10}  \dfrac{1}{c}
\]

\begin{example}
Let us consider an example of an idealized reaction system of the following type:
\begin{center}
\begin{tabular}{ | c | l | c | c | c | c |}
	\hline
	№ &  Reaction scheme & $v^{+}$ & $E^{+}$ & $v^{-}$ & $E^{-}$\\
	\hline
	1 & $A + C \rightarrow B + D$ & 200 & 350 & 0 & 300 \\
	\hline
	2 & $B + C \rightarrow I_{1}$ & 25 & 250 & 0 & 210 \\
	\hline
	3 & $B + I_{1} \rightarrow 3 B$ & 2000 & 210 & 0 & 200 \\
	\hline
	4 & $D + E \rightarrow C + F$ & 80 & 300 & 0 & 340 \\
	\hline
	5 & $B + D \rightarrow I_{2} + D$ & 20 & 100 & 0 & 110 \\
	\hline
	6 & $I_{2} + F \rightarrow C + F$ & 2000 & 110 & 0 & 120 \\
	\hline
	7 & $C + F \rightarrow I_{3}$ & 20 & 200 & 0 & 230 \\
	\hline
	8 & $C + I_{3} \rightarrow 2B + E$ & 80 & 230 & 0 & 240 \\
	\hline
\end{tabular}
\end{center}
To be exact, let us set all densities $\rho_{p} = 1$ and heat capacities $c_{p}=1$ for all substances, and also fix $H=L=W=1$, $V=a^{3}= 5100$, $T(0) = 40$, $N_{A}=N_{C}=2000$ and $N_{E}=1000$.
The kinetic curves that we have obtained for this example are shown in Figure \ref{nazarov:NazarovM_example}.
\begin{figure}[hbt]
\centering
\includegraphics[width=0.65\textwidth]{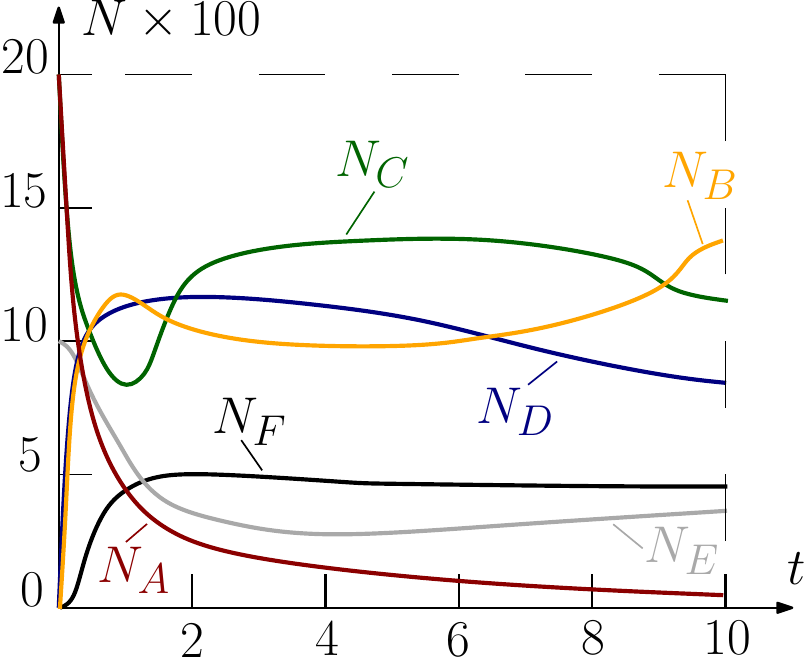}
\caption{ An example of quasi-periodic oscillations for a closed reaction system.}
\label{nazarov:NazarovM_example}
\end{figure}

An important feature of the reaction system under consideration is the quasi-periodic nature of the oscillations of the numbers $N_{C}$ and $N_{B}$ before reaching equilibrium. It should be noted that such oscillations are observed despite the fact that the system is closed and completely balanced in terms of the consumption and release of thermal energy. If we assume that the maximum concentration of substances is responsible for the color of the solution, then by choosing $H=1$ and $L=W=10$ we obtain the following picture ($D_{p} = 100$ for all substances):
\begin{figure}[hbt]
\centering
\includegraphics[width=0.8\textwidth]{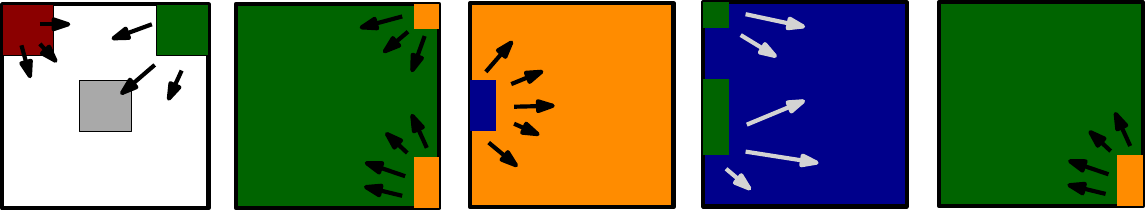}
\caption{Change in the dominant concentration in the reaction medium.}
\label{nazarov:NazarovM_3dex}
\end{figure}
\end{example} 
 
\begin{remark}
The main advantage of the approach considered is the simplicity  of its numerical implementation. However, it is not recommended to use such a model in problems where precise quantitative predictions are required (the accuracy estimate is given in \cite{Nazarov2014}).
\end{remark}



\section*{Conclusion}

Using the \eqref{nazarov:Chem_Kinetics_gener:main_system_eq1}--\eqref{nazarov:Delta_V_L_M_definition} model  one can, if desired, construct examples with complex configurations and dissipative structures in space (see examples in \cite{Vanag2004} and \cite{Rosov2011}). However, this would require allowing for exchange of matter with the external environment, as well as making the diffusion coefficients $D_{p}$ differ by an order of magnitude for different substances. For example, if catalysts are made less mobile, reaction zones will form around their clusters.

As has already been noted, the main advantage of the considered model \eqref{nazarov:Chem_Kinetics_gener:main_system_eq1}--\eqref{nazarov:Delta_V_L_M_definition} is the simplicity of its numerical implementation. But it's equally important to consider that using a system of ordinary differential equations instead of a partial differential system also allows for a qualitative analysis of dynamics using methods from dynamic systems theory. This allows for the study of both the overall steady states of the system and the partial steady states (introduced for a single cell or group of cells). Another important advantage is the ease of parallel implementation of the model  with the potential to distribute computations down to the level of individual cells.

Alternatively, instead of the numbers $N_{p}$, the model equations can use the amounts of substance $\nu_{p} = N_{p}/N_{A}$, the molar concentrations $[p] = \nu_{p} / a^{3} $, the molecular concentrations $[p^{N}] = N_{p} / a^{3} $, or the volume fractions $[p]_{d} = \nu_{p} /(a^{3}\rho^{\nu}_{p} ) = N_{p}/(a^{3}\rho^{N}_{p}) $. For example, to use $\nu_{p} = N_{p}/N_{A}$ it is enough to divide both parts of the equations \eqref{nazarov:Chem_Kinetics_gener:main_system_eq1} by the Avogadro's number $N_{A}$ and replace in the remaining equations $N_{p} = \nu_{p} \cdot N_{A} $. 

It is also possible to generalize the model to the case of a known laminar flow of matter in the space under consideration. Such a generalization is necessary for the model to be applicable to describing the dynamics of chemical reactions in which reactants mix at a constant rate. In fact, this requires transforming the known vector field defining the motion of matter into flows across cell boundaries.


\bibliographystyle{unsrt}  


\end{document}